\documentclass[10pt]{IETBook}

\begin{document}

\rhbooktitle{AI in Digitalization and Energy Management}

\markboth{AI in Digitalization and Energy Management}{AI in Digitalization and Energy Management}

\cauthor{
Jirawat Tangpanitanon,
\thanks{Quantum Technology Foundation (Thailand) Bangkok, Thailand}
\thanks{Thailand Center of Excellence in Physics, Ministry of Higher Education, Science, Research and Innovation, Bangkok, Thailand}
}

\setcounter{chapter}{20} 
\chapter{Quantum Computing for Energy Management: A Semi Non-Technical Guide for Practitioners} 

The pursuit of energy transition necessitates the coordination of several technologies, including more efficient and cost-effective distributed energy resources (DERs), smart grids, carbon capture, utilization, and storage (CCUS), energy-efficient technologies, Internet of Things (IoT), edge computing, artificial intellience (AI) and nuclear energy, among others. Quantum computing is an emerging paradigm for information processing at both hardware and software levels, by exploiting quantum mechanical properties to solve certain computational tasks exponentially faster than classical computers. This chapter will explore the opportunities and challenges of using quantum computing for energy management applications, enabling the more efficient and economically optimal integration of DERs such as solar PV rooftops, energy storage systems, electric vehicles (EVs), and EV charging stations into the grid. \footnote{This article is to be published by The Institution of Engineering and Technology, 2025.}


\section{Introduction}

Transitioning from fossil fuels due to climate change requires the adoption of green energy sources, leading to the rise of distributed energy resources (DERs) such as solar panels, wind turbines, energy storage systems, and electric vehicles \cite{akorede2010}. This energy transition is transforming the energy landscape and presenting new computational challenges, among others, to energy management due to the variability and decentralization of DERs \cite{abdmouleh2017, arturo2010}. These challenges include, for example, managing real-time optimal power flow to ensure grid stability and efficiency \cite{guerrero2020}, developing effective DER control strategies for reliable virtual power plant operations \cite{nosratabadi2017}, determining optimal locations for EV charging facilities \cite{tao2021}, and devising trading strategies for peer-to-peer energy markets \cite{zedan2024}. The underlying models frequently exhibit complex nonlinearities associated with power flow, memory effects from energy storage, a diverse array of DERs from local households to grid scales, and uncertainties stemming from the weather-dependent nature of renewable sources.

Quantum computing offers the potential to address these challenges with greater speed and efficiency, surpassing the capabilities of traditional computational approaches \cite{nielsen2010}. Unlike a classical computer that uses bits as the fundamental unit of computation, a quantum computer uses quantum bits, or `qubits', which can exist in a superposition of both 0 and 1 simultaneously. Multiple qubits can be `entangled' to produce correlations that cannot be reproduced by a classical computer. 

From computational complexity perspective, these non-classical properties enable a class of decision problems, called bounded-error quantum polynomial time (BQP) that can be efficiently solved by a quantum computer with certain error probability. BQP contains P, which is a class of decision problems efficiently solved by a classical computer or, more precisely, a deterministic Turing machine. In non-technical terms, this means that fundamentally there exists certain computational problems that can be efficiently solved by a quantum computer but not by any imaginable classical computers — a capability referred to as `quantum advantage' \cite{dominik2023}.

\begin{figure*}
\includegraphics[scale=0.6]{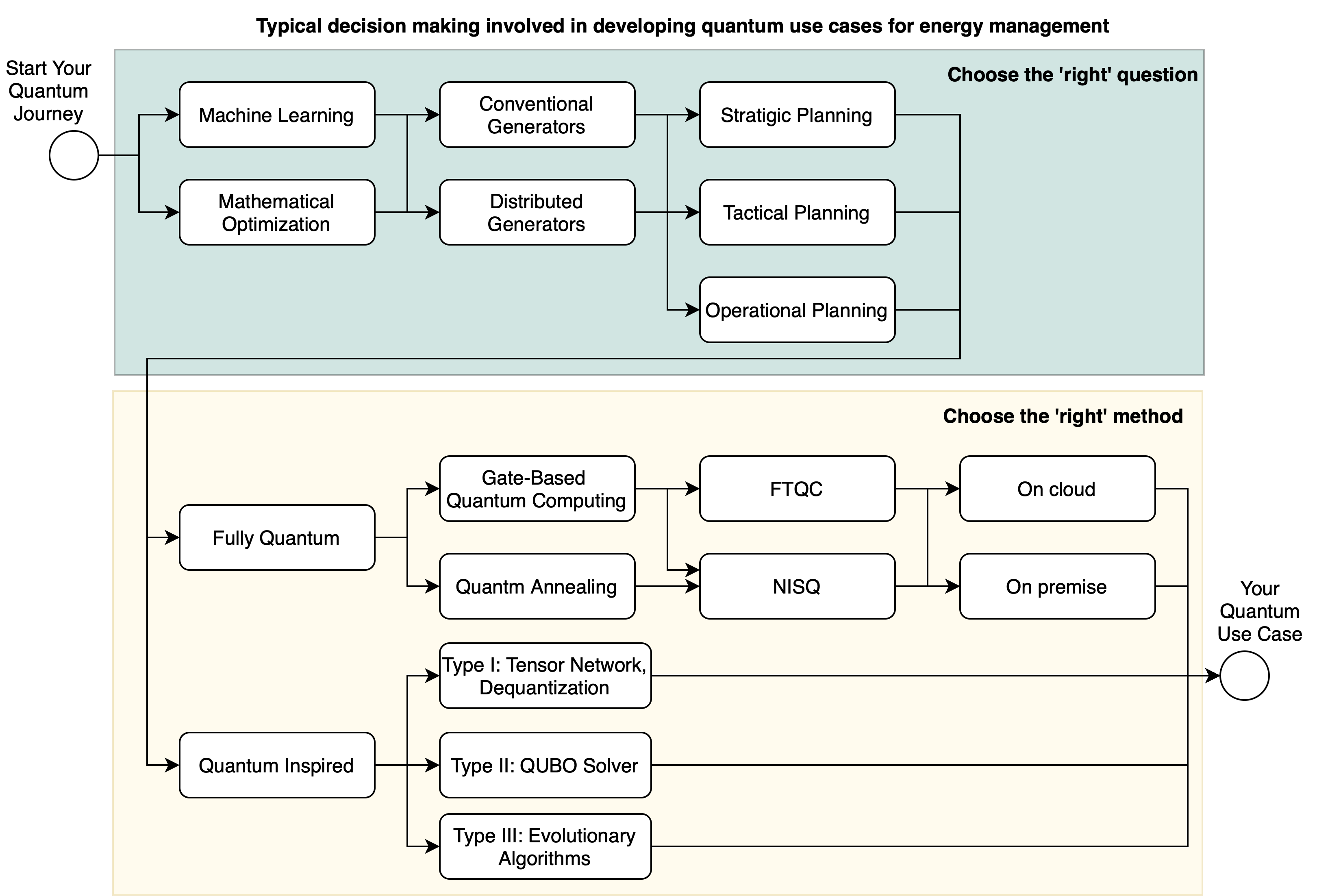}
\caption{A diagram illustrating the typical decision-making process for developing quantum computing use cases in energy management. Each option presents various pros and cons concerning immediate impacts, provable quantum speedup, practicality, cost, and learning opportunities.}
\label{fig1}
\end{figure*}

In the four decades since its inception by Richard Feynman in 1982 \cite{feynman1982}, quantum computing has achieved several breakthroughs \cite{preskill2023}, including advancements in its theoretical foundations, Nobel Prize-winning experiments, and the demonstration of quantum advantage in 2019 \cite{arute2019}. Recently, quantum computing had expanded from purely academic research to commercial ventures \cite{macquarrie2020}, with global investments from industry and government exceeding 30 billion USD in 2022 \cite{baloo2022}. Leading cloud providers such as AWS, Microsoft, and IBM offer access to quantum computers through a Software-as-a-Service (SaaS) model, enabling practitioners to run quantum experiments on actual quantum systems ranging from five to hundreds of qubits \cite{golec2024}. This `democratization' of quantum technology has spurred numerous studies on its real-world applications, ranging from artificial intelligence and optimization to simulations in various industry sectors, including finance \cite{nguyen2024, herman2022, daniel2020, orus2019}, energy \cite{ajagekar2019, ajagekar2022, giani2021, paudel2022, golestan2023,morstyn2024, yifan2022, ullah2022, kenta2023, eskandarpour2020review}, materials science \cite{bauer2020}, logistics \cite{stuart2021, eneko2022}, aviation \cite{pontus2020}, supply chain management \cite{hassan2024}, telecommunications \cite{frank2023}, insurance \cite{muhsin2023}, and the pharmaceutical industry \cite{zinner2022}, among others.

Despite these advancements, it is important to note that there is currently no experimental demonstration of quantum advantage in solving real-world problems. The quantum advantage experiment conducted by the Google Quantum AI team in 2019 used specific mathematical problems that maximally favored quantum computation—a task of simulating quantum computation itself with randomized evolution \cite{arute2019}. Additionally, current quantum hardware is `noisy', limiting the number of logical operations that can be performed before quantum information is destroyed by a process called decoherence. These noisy qubits are referred to as physical qubits and `stable', which are error-corrected and more reliable for computations, qubits are called logical qubits \cite{devitt2013}. A few tens of logical qubits have recently been realized experimentally \cite{dolev2024}. However, depending on the applications, real-world use cases of quantum computing may require at least hundreds or thousands of high-quality logical qubits which may still take years to realize \cite{amara2024, isaac2022}. 

It is a biased intention of this chapter not to leave the readers with an `interesting but not now' impression of quantum computing. First, although a fully-fledged large-scale quantum computer is still far fetching, energy transition is also far fetching involving strategic planning over several decades \cite{chen2019}. Ignoring advances in emerging technologies like quantum computing could lead to over-investment in outdated strategies, resulting in higher electricity costs or even unnecessarily unfeasible business models. Second, a quantum computer is not an advanced digital computer. Integrating a quantum computer into existing digital infrastructures is a long term process which requires multi-discipline collaborations including quantum experts, system operators, power engineers, software engineers, network engineers, business developers, policy makers and regulators \cite{kaur2022}. Third, there exists `quantum-inspired algorithms', such as tensor networks, that can be run on an existing classical computer by approximating quantum computation. Applications of these techniques for improving classical machine learning and optimization algorithms have been demonstrated \cite{cichocki2017, huggins2019}. Forth, there are `quantum-inspired hardware' such as Fujitsu's digital annealer \cite{satoshi2020} and Toshiba's bifurcation machine \cite{goto2016} that are currently capable of solving large-scale optimization problems. These are classical hardware designed to mimic certain aspects of quantum optimization.  

There is a wealth of material introducing quantum computing, along with an extensive list of prominent quantum algorithms \cite{abhijith2022, ashley2016, dalzell2023}. Additionally, several reviews discuss the applications of quantum computing in energy management \cite{ajagekar2019, ajagekar2022, giani2021, paudel2022, golestan2023,morstyn2024, yifan2022, kenta2023, eskandarpour2020review, ullah2022}. Instead of reiterating this existing content, this chapter intends to offer practical guidelines for energy practitioners interested in developing their own quantum computing use cases. We will begin by discussing the typical decision-making process involved in defining the research question and methodology for various quantum solutions. The chapter's layout is summarized in Fig. \ref{fig1}. Note that the references mentioned serve as examples and do not necessarily encompass the entirety of the existing literature.


\section{Choosing the `right' questions}
\label{sec:questions}

An extensive literature on quantum computing use cases in the energy sector showcases a broad range of applications and involves various quantum methods at different stages of maturity. For first-time explorers, finding applications that align with their interests can be overwhelming. While there is no definitive right or wrong approach to selecting quantum computing use cases, this section aims to outline the pros and cons of various decision-making choices related to typical questions that may arise during the process.

\subsection{Machine learning or mathematical optimization}

Machine learning (ML) and mathematical optimization (MO) are two distinct but interconnected paradigms within the field of data analytics. ML leverages large datasets to train models that identify patterns and make predictions about unknown data. Conversely, MO utilizes relatively smaller datasets to define business logic and applies search algorithms to find optimal decisions that minimize a cost function, subject to a set of constraints. 

There is an extensive body of literature on the applications of ML and MO in energy management \cite{chang2021, perera2021}, ranging from forecasting of variable renewable energy (VRE) production \cite{abualigah2022} to solving complex optimal power flow \cite{yujie2017}. To understand the new possibilities that quantum computing brings to these applications, it is beneficial to examine the theoretical foundations of quantum speedup in ML and MO.

\subsubsection{Quantum machine learning}

In the realm of machine learning, several quantum machine learning (QML) models \cite{cerezo2022, biamonte2017, schuld2014} are expected to have exponential speedup compared to their classical counterparts, including least squares fitting \cite{wiebe2012}, quantum Boltzmann machines \cite{mohammad2018}, quantum principal component analysis \cite{lloyd2014} and quantum support vector machines \cite{rebentrost2014}. Additionally, there are models that demonstrate quadratic speedup, such as Bayesian inference \cite{hao2014}, online perception \cite{ashish2016}, and quantum reinforcement learning \cite{vedran2016}. These speedups are achieved by substituting classical subroutines with more efficient quantum alternatives, such as matrix inversion, estimation of eigenvectors and eigenvalues for sparse or low-rank matrices, and fast Fourier transforms. These quantum subroutines frequently rely on the quantum Fourier transform (QFT) \cite{coppersmith2002} and the Harrow–Hassidim–Lloyd (HHL) algorithm \cite{aram2009}, the latter being a quantum method for numerically solving systems of linear equations. 

In the energy domain, the HHL algorithm has been explored as a solution for the model predictive control (MPC) problem in nonlinear dynamical microgrids \cite{jing2024}. Additionally, several studies have applied the HHL algorithm to address power flow issues, which are common nonlinear challenges encountered in energy grids \cite{feng2023, sævarsson2022, eskandarpour2021,feng2021, eskandarpour2020}. 

There are a few caveats regarding the QML models mentioned above. First, unless working with very small datasets, they require large-scale fault-tolerant quantum computers, which are not yet available. Second, these models necessitate the use of Quantum Random Access Memory (QRAM) \cite{vittorio2008}, potentially requiring an exponential amount of quantum resources to load classical data into a quantum computer. Exploring end-to-end advantages of QML models over their classical counterparts remains an active area of research.

Additionally, there is a relatively new class of quantum algorithms called variational quantum algorithms (VQAs) that can be used for machine learning \cite{cerezo2021}. Unlike QFT and HHL, VQAs involve variational parameters. These parameters are optimized using a classical computer based on the results from multiple runs of a quantum computer. Consequently, VQAs are more robust to noise. This makes them more compatible with noisy intermediate-scale quantum (NISQ) devices \cite{preskill2018}, which refer to current quantum hardware consisting of several hundreds or thousands of physical qubits. 

VQA has been used for photovoltaic power forecasting \cite{jeong2024, sagingalieva2023, munim2023} as well as solar irradiance \cite{oliveirasantos2024}. The latter uses the Folsom dataset, which spans from 2014 to 2016, containing minute-by-minute measurements of Global Horizontal Irradiance and Direct Normal Irradiance from Folsom, California, USA. Additionally, this method has been applied to forecast smart grid stability using a small dataset consisting of four nodes \cite{batuhan2024}.

One of the drawbacks of this approach is the scalability of VQAs. Due to the nature of the quantum parameter space, the optimization landscape observed by a classical optimizer often becomes exponentially flat with an increasing number of variational parameters—a phenomenon known as barren plateaus \cite{lennart2021, mcclean2018}. This problem can be mitigated, to certain extend, with good initial guess.

Lastly, we also want to mention a more modern quantum kernel method \cite{schuld2021}. The idea behind quantum kernel methods is to use quantum computers to compute the kernel function for data embedded in a quantum feature space. The quantum feature space can represent data in a more complex and richer way than classical approaches, potentially leading to better performance in tasks like classification or regression.

Currently, there are theoretical foundation proving quantum advantage for QML, but this is limited to specific tasks, such as when the learning data originates from quantum systems, such as those in quantum chemistry \cite{gyurik2023, molteni2024} and cryptography problems \cite{liu2021}. However, practical applications related to these tasks remain unclear.

Based on this observation, there is currently no apparent, at least to the author, quantum speedup for machine learning applications in energy management. However, it's important to clarify that we refer to quantum speedup in a strong sense, where a QML model must outperform any known classical ML models. In practice, it's possible to achieve a `quantum-enhanced' ML model that outperforms a specific classical model in a particular use case. Much like the effectiveness of machine learning techniques such as deep learning, which often emerges from empirical results rather than purely theoretical proofs, QML might similarly show advantages in the future. Given that large-scale quantum computers do not yet exist, the current lack of theoretical foundation does not necessarily preclude future advantages over classical ML.

\subsubsection{Quantum optimization}

Mathematical optimization has been central to Operations Research (OR) since World War II \cite{hillier2015}. Finding the best decision variables for real-world optimization problems often falls into the NP-hard class of decision problems. It is a strongly-held conjecture in computer science that no classical algorithms can efficiently solve NP-hard problems. Unfortunately, the same limitation applies to quantum computers because the complexity class BQP does not include or overlap with NP-hard problems. Nevertheless, there are certain classes of optimization problems whose optimal solutions can be efficiently `approximated' by a quantum algorithm, but not by any known classical algorithms \cite{pirnay2024, jordan2024}. Identifying these problems and exploring their real-world applications is currently an active area of research \cite{abbas2024}.

One of the most common formulations used in quantum optimization is Quadratic Unconstrained Binary Optimization (QUBO) problems \cite{glover2019}. The task is to find a set of $N$ binary variables $x_i \in \{0,1\}$ that minimizes an objective function 
\begin{equation}
\mathcal{C}_{\text{QUBO}} = \sum_{i,j=1}^N c_{i,j} x_i x_j,
\end{equation}
where $c_{i,j} \in \mathbb{R}$ is a QUBO coefficient. QUBO is computationally equivalent to the Ising model, which is commonly used to model spin systems in physics.

It is important to note that the widespread usage of QUBO in quantum optimization is primarily driven by the nature of quantum hardware itself, rather than its direct application to real-world problems. The QUBO formalism displays several distinct features compared to the commonly used Mixed-Integer Linear Programming (MILP). MILP involves finding a set of decision variables $u_i \in \mathbb{Z}^+$ and $u'_j \in \mathbb{R}^+$ to minimize a linear objective function
\begin{equation}
\mathcal{C}_{\text{MILP}} = \sum_{i=1}^{N_u} c_i u_i + \sum_{j=1}^{N_{u'}} c'_j u'_j,
\end{equation}
subject to linear constraints
\begin{equation}
\sum_{i=1}^{N_u} a_{k,i} u_i + \sum_{j=1}^{N_{u'}} a'_{k,j} u'_j \leq b_k,
\end{equation}
where $k \in \{1, 2, \ldots, K\}$, and $N_u, N_{u'}, K \in \mathbb{Z}^+$, with $c_i, c'_j, a_{k,i}, a'_{k,j}, b_k \in \mathbb{R}$.

There exists a standardized protocol for transforming MILP into QUBO. First, it is necessary to represent $u_i$ and $u'_j$ in binary form by introducing additional binary variables, analogous to the floating-point representation of real numbers \cite{braine2021}. Next, all `hard' constraints in the MILP are to be converted into `soft' constraints, or equivalently, into a quadratic cost function within the QUBO framework, through the introduction of a Lagrange multiplier \cite{djidjev2023}.

Example works that utilize the above transformation include:
\begin{itemize}
    \item The optimal placement of phasor measurement units (PMUs), aimed at minimizing the number of PMUs required while ensuring complete network observability according to established observability rules \cite{jones2020}.
    \item The placement of electric vehicle charging stations, which focuses on identifying the combination of locations that requires the fewest charging stations \cite{radvand2024, rao2023, margarita2023, rozycki2023, aman2022}.
    \item The optimal scheduling of battery energy storage systems to minimize total operating costs while adhering to charging and discharging constraints \cite{ehsani2024, kea2023}.
    \item The formulation of optimal charging schedules for electric vehicles in conjunction with local solar power generation, aiming to minimize reliance on the power grid \cite{marika2022}.
    \item The optimal discount scheduling problems for demand response, a strategy that enables consumers to actively manage electricity demand \cite{bucher2024}.
\end{itemize}

There are a few caveats associated with the aforementioned transformation. First, converting hard constraints into soft constraints can lead to solutions that are optimal within the QUBO framework but infeasible in the original MILP framework. Second, the quadratic cost function related to the soft constraints can become highly complex, involving all-to-all connectivity among decision variables. This complexity can transform a relatively simple MILP problem into a significantly more complex QUBO problem. Lastly, the number of QUBO coefficients increases quadratically with the number of decision variables. Consequently, a medium-sized MILP with, for instance, $10^5$ variables may transform into a gigantic QUBO instance with more than $10^{10}$ coefficients. This large number of coefficients could pose challenges related to communication bandwidth when attempting to send the data to a quantum computer.

Instead of transforming MILP to QUBO, it should be noted that it is possible, and should be encouraged, to devise a problem that is fundamentally QUBO. A prominent example of such a problem is the quadratic assignment problem (QAP). In QAP, there are a collection of $N$ facilities and $N$ locations. For each pair of locations $(i,i')$, a distance $d_{i,i'}$ is specified and for each pair of facilities $(j,j')$ a weight or flow $w_{j,j'}$ is specified (e.g., the amount of supplies transported between the two facilities). The problem is to assign all facilities to different locations with the goal of minimizing the sum of the distances multiplied by the corresponding flows. The objective function takes the form
\begin{equation}
\mathcal{C}_{\text{QAP}} = \sum_{i,i',j,j'=1}^{N} d_{i,i'} w_{j,j'} x_{i,j}x_{i',j'},
\end{equation}
where $x_{i,j}$ ($x_{i',j'}$) is equal to one if location $i$ ($i'$) is assigned with facility $j$ ($j'$) and zero otherwise. Facility locations allocation problems of power plants have been studied with this approach \cite{ajagekar2019}. A related problem has been studied for the tactical capacity problem to minimize loss in distribution networks \cite{vanderLinde2023, silva2023}.

Another standard QUBO problem is a maximum cut problem. Here, the objective is to partition the graph's vertices into two complementary sets $S$ and $T$, such that the number of edges between $S$ and $T$ is as large as possible. The corresponding cost function is
\begin{equation}  
\mathcal{C}_{\text{MaxCut}} = \sum_{(u,v)\in E} \frac{1}{2}(1-x_ux_v),
\end{equation}
where $E$ is a set of edges and $x_u$ ($x_v$) denotes whether node $u$ ($v$) is in subset $S$ or $T$. This formulation has been used to analyze stability of the power grid \cite{kaseb2024, bauer2024, bucher2024evaluatingquantumoptimizationdynamic, hang2022}.

A typical quantum approach to solving QUBO problems involves using either a specific type of VQAs known as Quantum Approximate Optimization Algorithms (QAOA) \cite{jaeho2019}, or another class of quantum methods known as quantum annealing \cite{yarkoni2022}. Although these two methods are theoretically related and do not require a fault tolerant quantum computer, they require different types of quantum hardware which will not be discussed in this chapter. Similar to machine learning applications, in both cases, there is currently no theoretical guarantee of quantum speedup \cite{abbas2024}. Again, this lack of theoretical foundation does not preclude advantage of quantum optimization in solving particular real-world problems.  

Lastly, it is worth highlighting recent proofs showing that quantum computers can achieve a super-polynomial advantage over classical computers for approximating certain combinatorial optimization problems. The problems addressed include the formula coloring problem \cite{pirnay2024}, optimal polynomial intersection \cite{jordan2024}, and max-XORSAT \cite{jordan2024}, rather than QUBO problems. This advantage utilizes the well-known Shor's and QFT algorithms, which require a fault-tolerant quantum computer. Mapping these problems to energy management applications presents an interesting avenue for further research.

\subsection{Conventional or distributed generators}

When selecting use cases for quantum computing in energy management, it’s essential to consider whether the focus is on conventional or distributed generators. Common optimization challenges for conventional generators include, for example, unit commitment (UC) \cite{saravanan2013} and economic dispatch (ED) \cite{xia2010}. UC involves determining which power generation units should be operational at any given time to meet anticipated electricity demand while adhering to operational constraints. It effectively decides the ‘commitment’ of generators for the following day or operational period. ED, on the other hand, is concerned with optimizing the output levels of the committed generators to minimize overall generation costs, while also meeting demand and complying with operational constraints.

The application of quantum methods to UC and ED presents the advantage of addressing well-defined problems with clear business impacts \cite{ajagekar2019, samantha2021, nima2022, reza2023}. However, there are significant challenges. System operators have often refined their current methods over decades, making it highly likely that these approaches are already near optimum, leaving little room for quantum computing to yield improvements. Additionally, regulatory frameworks in some regions impose specific planning guidelines that must be followed. For instance, ‘must-run’ power plants are prioritized to ensure grid stability, followed by ‘must-take’ plants based on long-term power purchase agreements. After these, remaining units are dispatched in merit order, giving preference to the most cost-effective options. Such regulations further streamline the computational challenges in optimizing generation strategies.

Distributed generators (DG), such as battery stations, solar rooftops, and very-small power producers, offer a distinct set of advantages and challenges for quantum computing applications compared to conventional generators. DG can be deployed in various contexts and introduces new computational challenges, particularly in managing distribution grid stability, which often involves non-linear optimization \cite{abdmouleh2017, arturo2010}. As a result, DG presents a broader range of use cases for exploring quantum computing applications. However, in many countries, DG is still emerging and faces regulatory, management, and economic challenges. Consequently, the immediate impact of quantum computing on DG may not be readily apparent.

\subsection{Strategic, tactical, or operational}

Another important consideration is the phase of operations related to quantum computing use cases: strategic, tactical, and operational. Each phase presents its own set of potential challenges and benefits associated with the implementation of quantum computing. 

\subsubsection{Strategic phase}

The strategic phase involves long-term planning for infrastructure investments, particularly in the energy sector, where plans often extend over several decades. On a national scale, these plans might include integrating more renewable energy sources such as wind, solar, and hydroelectric power to reduce carbon emissions and reliance on fossil fuels. They may also involve constructing and enhancing energy infrastructure, such as power plants, transmission lines, and smart grid technologies, to meet future demands and improve reliability \cite{terrados2007}. On an enterprise scale, manufacturing companies might focus on strategically planning their renewable energy portfolio, Battery Energy Storage Systems (BESS), and EV fleets to align with global efforts to reduce carbon emissions \cite{seetharaman2016}. 

The strategic phase presents an interesting opportunity for quantum optimization. Firstly, strategic planning usually involves large MO models that may take standard MILP solvers hours or even days to solve to yield satisfactory solutions. Therefore, there is a gap that quantum optimization can attempt to fill. Secondly, slight improvements in strategic planning can have significant financial impacts or even make previously infeasible business models feasible, justifying the investment in quantum computing. Thirdly, since strategic planning is typically a non-recurring activity, the cost of quantum computing can be viewed as a one-time capital expenditure (CAPEX) investment rather than an operational expense (OPEX), as it is in other phases.

However, a drawback of employing quantum computing in this phase is that the underlying MO models may not fully encapsulate all the decisions and concerns expressed by all relevant stakeholders. These may include subjective decision making, qualitative data sources, business negotiations, environmental complaints, and unforeseen regulatory restrictions. Such factors can complicate the quantification of the benefits of quantum optimization in terms of actual business impacts.

\subsubsection{Tactical phase}

The tactical phase in energy management typically involves day-ahead planning. Examples include
\begin{itemize}
\item For system operators, a unit commitment problem has to be solved in order to schedule and dispatch electric power generation resources \cite{saravanan2013, xia2010}. 
\item For power plants, operators must provide system operators with specific information about their plant's capabilities for the next day to ensure reliable grid operation. This includes data such as maximum output capacity, minimum operating level, ramp rates, and available ancillary services. While these tasks are routine for traditional power plants, determining optimal daily commitments can be complex for virtual power plants \cite{morteza2016}. These facilities must coordinate their member DERs, which may be located in various locations and independently managed by different entities. 
\item In a peer-to-peer energy market, participants, including prosumers with their own generation capabilities, need to provide specific information for the next-day trading in a manner that optimizes their own objectives. This information typically includes energy generation forecasts, energy consumption forecasts, available excess energy, pricing information, trading preferences, and availability for ancillary services \cite{chen2019p2p}.
\item For EV-fleet operators, tactical planning may involve optimizing routes and charging schedules \cite{yang2015}. This may include sourcing or bidding for green energy from DERs to minimize costs and environmental impact. 
\item For manufacturing operators, tactical planning may involve coordinating machine scheduling with battery usage to effectively utilize solar energy generated from photovoltaic (PV) rooftop installations. This approach can help optimize grid loads for peak shaving or participate in demand response programs, improving efficiency and reducing energy costs \cite{fang2011}.
\end{itemize}
As seen above, the energy transition creates multiple optimization and forecasting opportunities for various players, from grid operators to prosumers. This makes it a compelling research area to explore quantum methods for addressing these challenges. Although tactical operations may require daily access to quantum resources, its OPEX remain relatively manageable, compared to the demands of real-time operations in the operational phase.

Quantum computing use cases in the tactical phase face a similar drawback to those in the strategic phase. Since actual operations often differ from forecasts, the potential speedup provided by quantum computing in the tactical phase might be compromised by these discrepancies. Developing robust MO models can help mitigate this drawback.

\subsubsection{Operational phase}

The operational phase is where actual business value is realized. However, during real-time operations, decision-making options are often constrained by physical and managerial limitations of ongoing activities. It is not immediately clear, at least to the author, whether using a quantum computer to explore a vast solution space in that situation would outperform a simpler, more frequent feedback control loop running heuristic algorithms on a classical computer, such as those proposed in Ref. \cite{anese2018} for virtual power plant control. Moreover, the operational phase's demand for continuous use of quantum resources could significantly increase OPEX that outweigh potential benefits. Balancing the advantages of quantum computing with its cost implications and practical feasibility in real-time applications remains a largely unexplored territory. 

In summary, each of the three phases—strategic, tactical, and operational—has its own pros and cons concerning quantum computing use cases. It's important to note that many of the challenges discussed arise not directly from quantum computing itself but from the fact that MO and ML models are only approximate representations of reality. Therefore, for quantum initiatives to succeed, it is crucial to incorporate expertise from various fields, including energy, quantum computing, artificial intelligence, and operations research.


\section{Choosing the `right' methods}
\label{sec:methods}

In the previous section, we discussed the classifications of quantum computing use cases for energy management based on analytical techniques and operational phases. In this section, we will explore key considerations for selecting the appropriate quantum methods for specific use cases.

\subsection{Quantum or quantum-inspired}

Quantum methods, also known as fully quantum methods, utilize quantum hardware for execution. In contrast, quantum-inspired methods leverage digital hardware to run software influenced by quantum principles. Only fully quantum methods can achieve quantum advantage in a strong sense, as defined by computational complexity theory. However, due to the current unavailability of large-scale, fault-tolerant quantum computers, pursuing quantum-inspired methods presents a viable alternative when seeking tangible improvements over existing solutions.

The exploration of fully quantum methods offers benefits by deepening practitioners' understanding of quantum computing from both hardware and software perspectives. As noted previously, a quantum computer is not an advanced digital computer. For instance, while digital computers operate on deterministic principles, quantum computers are inherently probabilistic. This means that running the same software on a quantum computer may yield different results each time. Grasping these fundamental differences is essential for formulating a long-term quantum readiness strategy. 

There are two primary types of fully quantum methods: gate-based quantum computing \cite{nielsen2010} and quantum annealing \cite{yarkoni2022}. A gate-based quantum computer is universal, meaning it can execute any quantum algorithm. In contrast, a quantum annealer is specifically designed to tackle a particular class of optimization problems, such as QUBO. Quantum advantage has only been demonstrated for gate-based quantum computers. Conversely, while a quantum annealer utilizes heuristic approaches, it does not require error correction making it more suitable for near-term applications. 

It is important to note that this classification is simplified. For instance, there are analog quantum computers that utilize continuous quantum states and employ continuous transformations instead of discrete qubits and gates. This type of quantum hardware is also commercially available. However, the details of analog quantum computing are beyond the scope of this chapter.

Quantum-inspired methods, on the other hand, can be categorized into three main types:
\begin{itemize}
\item Type I: This category employs tensor networks (TNs), numerical techniques originally developed to investigate the dynamics and ground state properties of quantum many-body systems \cite{biamonte2017tn}. Recent studies have noted similarities between certain tensor networks and artificial neural networks (ANNs) \cite{wang2023tn}. While most ANN models are non-linear, TNs are linear, making them comparatively easier to analyze. For instance, Singular Value Decomposition (SVD) can help clarify the role of long-range correlations in the prediction accuracy of TN models \cite{tangpanitanon2022}. The explainability of machine learning models is crucial, as it enables users to understand decision-making processes, fostering trust and accountability in real-world applications. TNs have recently been explored across both ML \cite{huggins2019} and MO \cite{cichocki2017} domains. Unrelated to TNs, we also like to mention dequantization techniques that can be used to classically approximate HHL-based algorithms \cite{tang2019, arrazola2020}. However, the details of which may be challenging for beginners. 
\item Type II: This type encompasses software solutions such as Fujitsu's digital annealer \cite{satoshi2020} and Toshiba's simulated bifurcation machines \cite{goto2016}. These solutions address QUBO problems using heuristic techniques that draw loosely from the principles of quantum annealing. Additionally, it is worth noting that the standard Gurobi solver also offers a QUBO solver option. This class of quantum-inspired solutions is particularly effective for exploring large-scale QUBO applications involving between $10^4$ and $10^5$ variables.
\item Type III: This category includes heuristic search algorithms \cite{gharehchopogh2023}, such as quantum-inspired genetic algorithms \cite{narayanan1996} and quantum-inspired particle swarm algorithms \cite{ke2010}. These methodologies introduce randomness to enhance exploitation through quantum superpositions. Although this type is simpler to implement, it typically neglects quantum entanglement, a crucial factor for achieving quantum advantage.
\end{itemize}

In summary, the `quantumness' of the aforementioned methods can be ranked from most to least quantum as follows: fully quantum, quantum-inspired Type I, Type II, and Type III. Methods that are more aligned with quantum principles tend to be more challenging to implement but offer closer proximity to quantum advantage, whereas those that are easier to implement may drift further from these advantages.

\subsection{NISQ or FTQC}

NISQ stands for Near-Term Intermediate Scale Quantum computing \cite{preskill2018}, while FTQC refers to Fault-Tolerant Quantum Computing \cite{shor1996}. NISQ encompasses current or near-term quantum hardware with approximately a few hundred to a few thousand noisy qubits. In contrast, FTQC pertains to future large-scale quantum systems that utilize logical qubits.

Working with NISQ devices offers hands-on experience with fundamental quantum computing concepts, including superposition, entanglement, wave function collapse, and quantum gate decompositions. Practitioners can begin by using their laptops to simulate a small quantum computer with approximately 5-20 qubits. However, exactly simulating $N$ qubit systems generally requires $\mathcal{O}(2^N)$ classical resources to store quantum states and $\mathcal{O}(4^N)$ for quantum gates. Consequently, as the number of qubits increases, laptop simulation quickly becomes impractical.

Next, practitioners can leverage advanced quantum simulators offered by cloud services to benchmark quantum algorithms with 20-60 qubits. These simulators typically utilize tensor networks for better utilization of classical resources \cite{pan2024}. Ultimately, these algorithms can be executed on NISQ devices, enabling exploration of the noise characteristics of quantum hardware and understanding the current limitations in qubit connectivity.

FTQC, in contrast, provides a solid theoretical framework for achieving quantum speedup, which can guide the exploration of relevant applications. However, studying FTQC often demands extensive mathematical analysis, which may seem abstract to non-experts. Despite this, most FTQC algorithms are built upon a few foundational `building-block' quantum algorithms, making it not impossible for enthusiastic practitioners to understand. These foundational algorithms include Shor's algorithm for factorization, Grover's search algorithm, and the Quantum Fourier Transform \cite{nielsen2010}.

\subsection{On-cloud or on-premise}

Most quantum hardware providers offer cloud access to their quantum computers using pay-as-you-go models, which are popular due to their relatively accessible costs. However, certain energy management applications might be legally required to use on-premise facilities. For example, operations related to system operators are often connected to national security concerns, requiring cloud facilities to function within regulated areas.

Hosting quantum hardware demands a fundamentally different infrastructure than standard digital data centers. Therefore, practitioners should prepare a long-term plan for the installation and maintenance of quantum facilities. In addition, unlike most digital computers, which use CMOS (Complementary Metal-Oxide-Semiconductor) technology, quantum computing hardware is commercially available across various implementation platforms, including trapped ions \cite{monroe2021}, neutral atom arrays \cite{wintersperger2023}, photonic systems \cite{zhong2020}, and superconducting qubits \cite{morten2020}. Each platform has unique advantages and disadvantages and requires significantly different facilities. Detailed discussions on these platforms are beyond the scope of this chapter.


\section{Some practical remarks}
\label{sec:remarks}

In this section, we address practical concerns that often arise when undertaking quantum projects. Readers primarily interested in an overview of quantum computing use cases for energy management may choose to skip this section.

\subsubsection{Hardware availability}

Unlike standard digital clouds, where resources are highly available, quantum resources are scarce. Submitting a job to quantum cloud services often results in long queue times, ranging from a few seconds to several hours. Some quantum cloud providers offer a dedicated session option for an additional cost to mitigate this issue. However, when integrating quantum clouds into a production environment, it is advisable to have backup digital servers capable of executing the same tasks in case the quantum server becomes unexpectedly unavailable. This ensures continuity and reliability in operations while leveraging quantum computing capabilities.

\subsubsection{Hardware cost}

In pay-as-you-go quantum clouds, providers typically charge based on the run time of quantum hardware. However, for beginners, estimating this run time prior to execution can be challenging. Users often write quantum programs using high-level quantum programming languages, which must then be compiled for compatibility with specific quantum hardware. This compilation process can involve breaking long-range quantum gates into smaller gates to align with the qubit connectivity constraints of the hardware.

Consequently, the same quantum program may result in a vastly different number of quantum gates across different problem instances after compilation, leading to cost unpredictability from a software deployment perspective. While most providers offer automation for the compilation process, it remains vital to cross-check the number of gates and relevant parameters before submitting to a quantum cloud to prevent unexpected expenses.

\subsubsection{Communication overhead}

As mentioned earlier, transmitting large data sets to quantum clouds can become a bottleneck in certain applications. For example, machine learning applications that utilize HHL algorithms require QRAM, which typically requires an exponentially large number of quantum resources to load classical data into a quantum computer. 

In MO applications such as QUBO, the number of coefficients grows quadratically with the number of qubits, often resulting in several gigabytes of data in many real-world scenarios. Therefore, it is essential to be cautious when evaluating the end-to-end quantum speedup compared to classical methods, as the overhead associated with data transmission and loading can significantly impact performance assessments.

\subsubsection{Open-source quantum software}

It is generally advisable for software engineers not to modify software that is functioning well, as digital software does not degrade over time. However, this principle does not apply to quantum computing. As quantum hardware and quantum error correction protocols evolve rapidly, quantum software that is operational today may not work a year from now. In addition, some quantum hardware providers may deprecate compatibility with certain versions of open-source software on an annual basis. Therefore, it is crucial for practitioners to keep their quantum software up-to-date to ensure compatibility with ongoing developments in quantum hardware. 

\subsubsection{Debugging quantum program}

Debugging quantum software can be a significant challenge for practitioners due to the probabilistic nature of quantum computing. It is helpful to categorize uncertainties into two classes.

The first class involves uncertainties that arise from the principles of quantum physics. These uncertainties are inherent and unavoidable, even with perfectly built quantum hardware. In fact, they are essential for achieving quantum speedup and should be embraced rather than eliminated. However, they can complicate the debugging process, especially for beginners. To mitigate this, it is advisable to start with small problem instances and utilize a state vector simulator. This type of simulator focuses on the probability function rather than sampling from it, effectively ignoring quantum uncertainties and providing clearer insights during debugging.

The second class of uncertainty originates from the imperfections of quantum hardware itself. NISQ devices particularly struggle with this issue as they lack robust error correction compared to FTQC. Debugging quantum software in the presence of this type of uncertainty can benefit from benchmarking NISQ devices with simulators that account for quantum uncertainties, allowing practitioners to better understand and manage the imperfections of the hardware while developing their algorithms.


\section{Conclusion}

In conclusion, this chapter has outlined the opportunities and challenges presented by quantum computing within the realm of energy management. By leveraging the unique properties of quantum systems, practitioners can potentially optimize complex problems associated with distributed energy resources, smart grids, and overall energy efficiency. We have discussed various use cases, from quantum machine learning to optimization, and the implications for conventional and distributed energy generation. 

Selecting appropriate methodologies, whether fully quantum or quantum-inspired, is critical in navigating this evolving landscape. While quantum advantage have not yet been universally proven in practical applications, especially in energy management, the groundwork laid by quantum-inspired algorithms and techniques demonstrates potential pathways for near-term applications. 

Moreover, as our understanding of both quantum technology and its applications in energy systems continues to mature, strategic, tactical, and operational phases must be clearly defined to align quantum capabilities with real-world needs. This requires ongoing collaboration across disciplines, ensuring that quantum computing innovations are effectively integrated into existing energy infrastructure. The journey towards a quantum-enabled energy landscape is long and complex, but with the right approach, it can lead to transformative efficiencies and a more sustainable energy future.

\bibliography{paper0.3}

\begin{thebibliography}{100}

\bibitem{akorede2010}
Akorede MF, Hizam H, Pouresmaeil E.
\newblock Distributed energy resources and benefits to the environment.
\newblock Renewable and Sustainable Energy Reviews. 2010;14(2):724--734.
\newblock Available from: \url{https://www.sciencedirect.com/science/article/pii/S1364032109002561}.

\bibitem{abdmouleh2017}
Abdmouleh Z, Gastli A, Ben-Brahim L, et~al.
\newblock Review of optimization techniques applied for the integration of distributed generation from renewable energy sources.
\newblock Renewable Energy. 2017;113:266--280.
\newblock Available from: \url{https://www.sciencedirect.com/science/article/pii/S0960148117304822}.

\bibitem{arturo2010}
Alarcon-Rodriguez A, Ault G, Galloway S.
\newblock Multi-objective planning of distributed energy resources: A review of the state-of-the-art.
\newblock Renewable and Sustainable Energy Reviews. 2010;14(5):1353--1366.
\newblock Available from: \url{https://www.sciencedirect.com/science/article/pii/S1364032110000146}.

\bibitem{guerrero2020}
Guerrero J, Gebbran D, Mhanna S, et~al.
\newblock Towards a transactive energy system for integration of distributed energy resources: Home energy management, distributed optimal power flow, and peer-to-peer energy trading.
\newblock Renewable and Sustainable Energy Reviews. 2020;132:110000.
\newblock Available from: \url{https://www.sciencedirect.com/science/article/pii/S1364032120302914}.

\bibitem{nosratabadi2017}
Nosratabadi SM, Hooshmand RA, Gholipour E.
\newblock A comprehensive review on microgrid and virtual power plant concepts employed for distributed energy resources scheduling in power systems.
\newblock Renewable and Sustainable Energy Reviews. 2017;67:341--363.
\newblock Available from: \url{https://www.sciencedirect.com/science/article/pii/S1364032116305160}.

\bibitem{tao2021}
Tao Y, Huang M, Chen Y, et~al.
\newblock Review of optimized layout of electric vehicle charging infrastructures.
\newblock Journal of Central South University. 2021;28:3268--3278.
\newblock Available from: \url{https://doi.org/10.1007/s11771-021-4842-3}.

\bibitem{zedan2024}
Zedan M, Nour M, Shabib G, et~al.
\newblock Review of peer-to-peer energy trading: Advances and challenges.
\newblock e-Prime - Advances in Electrical Engineering, Electronics and Energy. 2024;10:100778.
\newblock Available from: \url{https://www.sciencedirect.com/science/article/pii/S2772671124003589}.

\bibitem{nielsen2010}
Nielsen MA, Chuang IL.
\newblock Quantum Computation and Quantum Information.
\newblock Cambridge University Press; 2010.

\bibitem{dominik2023}
Hangleiter D, Eisert J.
\newblock Computational advantage of quantum random sampling.
\newblock Rev Mod Phys. 2023 Jul;95:035001.
\newblock Available from: \url{https://link.aps.org/doi/10.1103/RevModPhys.95.035001}.

\bibitem{feynman1982}
Feynman RP.
\newblock Simulating physics with computers.
\newblock International Journal of Theoretical Physics. 1982;21(6-7):467--488.

\bibitem{preskill2023}
Preskill J.
\newblock Quantum Computing 40 Years Later.
\newblock 2nd ed. Feynman Lectures on Computation. CRC Press; 2023.

\bibitem{arute2019}
Arute F, Arya K, Babbush R, et~al.
\newblock Quantum Supremacy Using a Programmable Superconducting Processor.
\newblock Nature. 2019;574:505--510.
\newblock Available from: \url{https://www.nature.com/articles/s41586-019-1666-5#citeas}.

\bibitem{macquarrie2020}
MacQuarrie ER, Simon C, Simmons S, et~al.
\newblock The emerging commercial landscape of quantum computing.
\newblock Nature Reviews Physics. 2020;2:596--598.
\newblock Available from: \url{https://doi.org/10.1038/s42254-020-00247-5}.

\bibitem{baloo2022}
Jaya~Baloo ea.
\newblock State of Quantum Computing: Building a Quantum Economy.
\newblock World Economic Forum. 2022;Available from: \url{https://www.weforum.org/publications/state-of-quantum-computing-building-a-quantum-economy/}.

\bibitem{golec2024}
Golec M, Hatay ES, Golec M, et~al.
\newblock Quantum cloud computing: Trends and challenges.
\newblock Journal of Economy and Technology. 2024;2:190--199.
\newblock Available from: \url{https://www.sciencedirect.com/science/article/pii/S2949948824000271}.

\bibitem{nguyen2024}
Nguyen HT, Krishnan P, Krishnaswamy D, et~al.. Quantum Cloud Computing: A Review, Open Problems, and Future Directions; 2024.
\newblock Available from: \url{https://arxiv.org/abs/2404.11420}.

\bibitem{herman2022}
Herman D, Googin C, Liu X, et~al.. A Survey of Quantum Computing for Finance; 2022.
\newblock Available from: \url{https://arxiv.org/abs/2201.02773}.

\bibitem{daniel2020}
Egger DJ, Gambella C, Marecek J, et~al.
\newblock Quantum Computing for Finance: State-of-the-Art and Future Prospects.
\newblock IEEE Transactions on Quantum Engineering. 2020;1:1--24.

\bibitem{orus2019}
Orús R, Mugel S, Lizaso E.
\newblock Quantum computing for finance: Overview and prospects.
\newblock Reviews in Physics. 2019;4:100028.
\newblock Available from: \url{https://www.sciencedirect.com/science/article/pii/S2405428318300571}.

\bibitem{ajagekar2019}
Ajagekar A, You F.
\newblock Quantum computing for energy systems optimization: Challenges and opportunities.
\newblock Energy. 2019;179:76--89.
\newblock Available from: \url{https://www.sciencedirect.com/science/article/pii/S0360544219308254}.

\bibitem{ajagekar2022}
Ajagekar A, You F.
\newblock Quantum computing and quantum artificial intelligence for renewable and sustainable energy: A emerging prospect towards climate neutrality.
\newblock Renewable and Sustainable Energy Reviews. 2022;165:112493.
\newblock Available from: \url{https://www.sciencedirect.com/science/article/pii/S1364032122003975}.

\bibitem{giani2021}
Giani A, Eldredge Z.
\newblock Quantum Computing Opportunities in Renewable Energy.
\newblock SN Computer Science. 2021;2(393).

\bibitem{paudel2022}
Paudel HP, Syamlal M, Crawford SE, et~al.
\newblock Quantum Computing and Simulations for Energy Applications: Review and Perspective.
\newblock ACS Engineering Au. 2022;2(3):151--196.
\newblock Available from: \url{https://doi.org/10.1021/acsengineeringau.1c00033}.

\bibitem{golestan2023}
Golestan S, Habibi MR, {Mousazadeh Mousavi} SY, et~al.
\newblock Quantum computation in power systems: An overview of recent advances.
\newblock Energy Reports. 2023;9:584--596.
\newblock Available from: \url{https://www.sciencedirect.com/science/article/pii/S2352484722025720}.

\bibitem{morstyn2024}
Morstyn T, Wang X.
\newblock Opportunities for Quantum Computing within Net-Zero Power System Optimization.
\newblock Joule. 2024;8(6):1619--1640.
\newblock Available from: \url{https://www.cell.com/joule/fulltext/S2542-4351(24)00155-7}.

\bibitem{yifan2022}
Zhou Y, Tang Z, Nikmehr N, et~al.
\newblock Quantum computing in power systems.
\newblock iEnergy. 2022;1(2):170--187.

\bibitem{ullah2022}
Ullah M, Eskandarpour R, Zheng H, et~al.
\newblock Quantum computing for smart grid applications.
\newblock IET Generation, Transmission \& Distribution. 2022;16(23):4239--4257.

\bibitem{kenta2023}
Kirihara K, Imai H, Kuroda E, et~al.
\newblock Exploring Potential Applications of Ising Machines for Power System Operations.
\newblock IEEE Access. 2023;11:68004--68017.

\bibitem{eskandarpour2020review}
Eskandarpour R, Bahadur~Ghosh KJ, Khodaei A, et~al.
\newblock Quantum-Enhanced Grid of the Future: A Primer.
\newblock IEEE Access. 2020;8:188993--189002.

\bibitem{bauer2020}
Bauer B, Bravyi S, Motta M, et~al.
\newblock Quantum Algorithms for Quantum Chemistry and Quantum Materials Science.
\newblock Chemical Reviews. 2020;120(22):12685--12717.
\newblock Available from: \url{https://doi.org/10.1021/acs.chemrev.9b00829}.

\bibitem{stuart2021}
Harwood S, Gambella C, Trenev D, et~al.
\newblock Formulating and Solving Routing Problems on Quantum Computers.
\newblock IEEE Transactions on Quantum Engineering. 2021;2:1--17.

\bibitem{eneko2022}
Osaba E, Villar-Rodriguez E, Oregi I.
\newblock A Systematic Literature Review of Quantum Computing for Routing Problems.
\newblock IEEE Access. 2022;10:55805--55817.

\bibitem{pontus2020}
Vikst\aa{}l P, Gr\"onkvist M, Svensson M, et~al.
\newblock Applying the Quantum Approximate Optimization Algorithm to the Tail-Assignment Problem.
\newblock Phys Rev Appl. 2020 Sep;14:034009.
\newblock Available from: \url{https://link.aps.org/doi/10.1103/PhysRevApplied.14.034009}.

\bibitem{hassan2024}
Hassan A, Bhattacharya P, Dutta PK, et~al., editors.
\newblock Quantum Computing and Supply Chain Management: A New Era of Optimization.
\newblock IGI Global; 2024.

\bibitem{frank2023}
Phillipson F.
\newblock Quantum Computing in Telecommunication—A Survey.
\newblock Mathematics. 2023;11(15).
\newblock Available from: \url{https://www.mdpi.com/2227-7390/11/15/3423}.

\bibitem{muhsin2023}
Tamturk M. Quantum Computing in Insurance Capital Modelling; 2023.
\newblock Available from: \url{https://www.mdpi.com/2227-7390/11/3/658}.

\bibitem{zinner2022}
Zinner M, Dahlhausen F, Boehme P, et~al.
\newblock Toward the institutionalization of quantum computing in pharmaceutical research.
\newblock Drug Discovery Today. 2022;27(2):378--383.
\newblock Available from: \url{https://www.sciencedirect.com/science/article/pii/S135964462100444X}.

\bibitem{devitt2013}
Devitt SJ, Munro WJ, Nemoto K.
\newblock Quantum error correction for beginners.
\newblock Reports on Progress in Physics. 2013 jun;76(7):076001.
\newblock Available from: \url{https://dx.doi.org/10.1088/0034-4885/76/7/076001}.

\bibitem{dolev2024}
Bluvstein D, Evered SJ, Geim AA, et~al.
\newblock Logical quantum processor based on reconfigurable atom arrays.
\newblock Nature. 2024;626(7997):58--65.

\bibitem{amara2024}
Katabarwa A, Gratsea K, Caesura A, et~al.
\newblock Early Fault-Tolerant Quantum Computing.
\newblock PRX Quantum. 2024 Jun;5:020101.
\newblock Available from: \url{https://link.aps.org/doi/10.1103/PRXQuantum.5.020101}.

\bibitem{isaac2022}
Kim IH, Liu YH, Pallister S, et~al.
\newblock Fault-tolerant resource estimate for quantum chemical simulations: Case study on Li-ion battery electrolyte molecules.
\newblock Phys Rev Res. 2022 Apr;4:023019.
\newblock Available from: \url{https://link.aps.org/doi/10.1103/PhysRevResearch.4.023019}.

\bibitem{chen2019}
Chen B, Xiong R, Li H, et~al.
\newblock Pathways for sustainable energy transition.
\newblock Journal of Cleaner Production. 2019;228:1564--1571.
\newblock Available from: \url{https://www.sciencedirect.com/science/article/pii/S0959652619314738}.

\bibitem{kaur2022}
Kaur M, Venegas-Gomez A.
\newblock Defining the quantum workforce landscape: a review of global quantum education initiatives.
\newblock Optical Engineering. 2022;61(8):081806.

\bibitem{cichocki2017}
Cichocki A, Phan AH, Zhao Q, et~al.
\newblock Tensor Networks for Dimensionality Reduction and Large-scale Optimization: Part 2 Applications and Future Perspectives.
\newblock Foundations and Trends® in Machine Learning. 2017;9(6):431--673.
\newblock Available from: \url{http://dx.doi.org/10.1561/220000006}.

\bibitem{huggins2019}
Huggins W, Patil P, Mitchell B, et~al.
\newblock Towards quantum machine learning with tensor networks.
\newblock Quantum Science and Technology. 2019;4(2):024001.
\newblock Available from: \url{https://iopscience.iop.org/article/10.1088/2058-9565/aaea94/meta}.

\bibitem{satoshi2020}
Matsubara S, Takatsu M, Miyazawa T, et~al.
\newblock Digital Annealer for High-Speed Solving of Combinatorial optimization Problems and Its Applications.
\newblock In: 2020 25th Asia and South Pacific Design Automation Conference (ASP-DAC); 2020. p. 667--672.

\bibitem{goto2016}
Goto H.
\newblock Bifurcation-based adiabatic quantum computation with a nonlinear oscillator network: Toward quantum soft computing.
\newblock Scientific Reports. 2016;6:21686.

\bibitem{abhijith2022}
J A, Adedoyin A, Ambrosiano J, et~al.
\newblock Quantum Algorithm Implementations for Beginners.
\newblock ACM Transactions on Quantum Computing. 2022 Jul;3(4):1–92.
\newblock Available from: \url{http://dx.doi.org/10.1145/3517340}.

\bibitem{ashley2016}
Montanaro A.
\newblock Quantum algorithms: an overview.
\newblock npj Quantum Information. 2016;2(1):15023.
\newblock Available from: \url{https://doi.org/10.1038/npjqi.2015.23}.

\bibitem{dalzell2023}
Dalzell AM, McArdle S, Berta M, et~al.. Quantum algorithms: A survey of applications and end-to-end complexities; 2023.
\newblock Available from: \url{https://arxiv.org/abs/2310.03011}.

\bibitem{chang2021}
Chang M, Thellufsen JZ, Zakeri B, et~al.
\newblock Trends in tools and approaches for modelling the energy transition.
\newblock Applied Energy. 2021;290:116731.
\newblock Available from: \url{https://www.sciencedirect.com/science/article/pii/S0306261921002476}.

\bibitem{perera2021}
Perera ATD, Kamalaruban P.
\newblock Applications of reinforcement learning in energy systems.
\newblock Renewable and Sustainable Energy Reviews. 2021;137:110618.
\newblock Available from: \url{https://www.sciencedirect.com/science/article/pii/S1364032120309023}.

\bibitem{abualigah2022}
Abualigah L, Zitar R, Almotairi K, et~al.
\newblock Wind, Solar, and Photovoltaic Renewable Energy Systems with and without Energy Storage Optimization: A Survey of Advanced Machine Learning and Deep Learning Techniques.
\newblock Energies. 2022;15(2):578.

\bibitem{yujie2017}
Tang Y, Dvijotham K, Low S.
\newblock Real-Time Optimal Power Flow.
\newblock IEEE Transactions on Smart Grid. 2017;8(6):2963--2973.

\bibitem{cerezo2022}
Cerezo M, Verdon G, Huang H, et~al.
\newblock Challenges and opportunities in quantum machine learning.
\newblock Nature Computational Science. 2022;2:567--576.

\bibitem{biamonte2017}
Biamonte J, Wittek P, Pancotti N, et~al.
\newblock Quantum machine learning.
\newblock Nature. 2017;549:195--202.

\bibitem{schuld2014}
Schuld M, Sinayskiy I, Petruccione F.
\newblock An introduction to quantum machine learning.
\newblock Contemporary Physics. 2014;56(2):172--185.

\bibitem{wiebe2012}
Wiebe N, Braun D, Lloyd S.
\newblock Quantum algorithm for data fitting.
\newblock Physical review letters. 2012;109(5):050505.

\bibitem{mohammad2018}
Amin MH, Andriyash E, Rolfe J, et~al.
\newblock Quantum Boltzmann Machine.
\newblock Phys Rev X. 2018 May;8:021050.
\newblock Available from: \url{https://link.aps.org/doi/10.1103/PhysRevX.8.021050}.

\bibitem{lloyd2014}
Lloyd S, Mohseni M, Rebentrost P.
\newblock Quantum principal component analysis.
\newblock Nature Physics. 2014;10:631--633.

\bibitem{rebentrost2014}
Rebentrost P, Mohseni M, Lloyd S.
\newblock Quantum support vector machine for big data classification.
\newblock Physical review letters. 2014;113(13):130503.

\bibitem{hao2014}
Low GH, Yoder TJ, Chuang IL.
\newblock Quantum inference on Bayesian networks.
\newblock Phys Rev A. 2014 Jun;89:062315.
\newblock Available from: \url{https://link.aps.org/doi/10.1103/PhysRevA.89.062315}.

\bibitem{ashish2016}
Kapoor A, Wiebe N, Svore K.
\newblock Quantum Perceptron Models.
\newblock In: Lee D, Sugiyama M, Luxburg U, et~al., editors. Advances in Neural Information Processing Systems. vol.~29. Curran Associates, Inc.; 2016. Available from: \url{https://proceedings.neurips.cc/paper_files/paper/2016/file/d47268e9db2e9aa3827bba3afb7ff94a-Paper.pdf}.

\bibitem{vedran2016}
Dunjko V, Taylor JM, Briegel HJ.
\newblock Quantum-Enhanced Machine Learning.
\newblock Phys Rev Lett. 2016 Sep;117:130501.
\newblock Available from: \url{https://link.aps.org/doi/10.1103/PhysRevLett.117.130501}.

\bibitem{coppersmith2002}
Coppersmith D. An approximate Fourier transform useful in quantum factoring; 2002.
\newblock Available from: \url{https://arxiv.org/abs/quant-ph/0201067}.

\bibitem{aram2009}
Harrow AW, Hassidim A, Lloyd S.
\newblock Quantum Algorithm for Linear Systems of Equations.
\newblock Phys Rev Lett. 2009 Oct;103:150502.
\newblock Available from: \url{https://link.aps.org/doi/10.1103/PhysRevLett.103.150502}.

\bibitem{jing2024}
Jing H, Li Y, Brandsema MJ, et~al.
\newblock HHL algorithm with mapping function and enhanced sampling for model predictive control in microgrids.
\newblock Applied Energy. 2024;361:122878.
\newblock Available from: \url{https://www.sciencedirect.com/science/article/pii/S0306261924002617}.

\bibitem{feng2023}
Feng F, Zhou YF, Zhang P.
\newblock Noise-resilient quantum power flow.
\newblock iEnergy. 2023;2(1):63--70.

\bibitem{sævarsson2022}
Sævarsson B, Chatzivasileiadis S, Jóhannsson H, et~al.. Quantum Computing for Power Flow Algorithms: Testing on real Quantum Computers; 2022.
\newblock Available from: \url{https://arxiv.org/abs/2204.14028}.

\bibitem{eskandarpour2021}
Eskandarpour R, Ghosh K, Khodaei A, et~al.. Experimental Quantum Computing to Solve Network DC Power Flow Problem; 2021.
\newblock Available from: \url{https://arxiv.org/abs/2106.12032}.

\bibitem{feng2021}
Feng F, Zhou Y, Zhang P.
\newblock Quantum Power Flow.
\newblock IEEE Transactions on Power Systems. 2021;36(4):3810--3812.

\bibitem{eskandarpour2020}
Eskandarpour R, Ghosh K, Khodaei A, et~al.. Quantum Computing Solution of DC Power Flow; 2020.
\newblock Available from: \url{https://arxiv.org/abs/2010.02442}.

\bibitem{vittorio2008}
Giovannetti V, Lloyd S, Maccone L.
\newblock Quantum Random Access Memory.
\newblock Phys Rev Lett. 2008 Apr;100:160501.
\newblock Available from: \url{https://link.aps.org/doi/10.1103/PhysRevLett.100.160501}.

\bibitem{cerezo2021}
Cerezo M, Arrasmith A, Babbush R, et~al.
\newblock Variational Quantum Algorithms.
\newblock Nature Reviews Physics. 2021;3:625--644.

\bibitem{preskill2018}
Preskill J.
\newblock Quantum Computing in the NISQ Era and Beyond.
\newblock Quantum. 2018;2:79.

\bibitem{jeong2024}
Jeong SG, Do QV, Hwang WJ.
\newblock Short-term photovoltaic power forecasting based on hybrid quantum gated recurrent unit.
\newblock ICT Express. 2024;10(3):608--613.
\newblock Available from: \url{https://www.sciencedirect.com/science/article/pii/S2405959523001637}.

\bibitem{sagingalieva2023}
Sagingalieva A, Komornyik S, Senokosov A, et~al.. Photovoltaic power forecasting using quantum machine learning; 2023.
\newblock Available from: \url{https://arxiv.org/abs/2312.16379}.

\bibitem{munim2023}
Sushmit MM, Mahbubul IM.
\newblock Forecasting solar irradiance with hybrid classical–quantum models: A comprehensive evaluation of deep learning and quantum-enhanced techniques.
\newblock Energy Conversion and Management. 2023;294:117555.
\newblock Available from: \url{https://www.sciencedirect.com/science/article/pii/S0196890423009019}.

\bibitem{oliveirasantos2024}
Oliveira~Santos V, Marinho FP, Costa~Rocha PA, et~al.
\newblock Application of Quantum Neural Network for Solar Irradiance Forecasting: A Case Study Using the Folsom Dataset, California.
\newblock Energies. 2024;17:3580.

\bibitem{batuhan2024}
Hangun B, Eyecioglu O, Altun O.
\newblock Quantum Computing Approach to Smart Grid Stability Forecasting.
\newblock In: 2024 12th International Conference on Smart Grid (icSmartGrid); 2024. p. 840--843.

\bibitem{lennart2021}
Bittel L, Kliesch M.
\newblock Training Variational Quantum Algorithms Is NP-Hard.
\newblock Phys Rev Lett. 2021 Sep;127:120502.
\newblock Available from: \url{https://link.aps.org/doi/10.1103/PhysRevLett.127.120502}.

\bibitem{mcclean2018}
McClean JR, Boixo S, Smelyanskiy VN, et~al.
\newblock Barren Plateaus in Quantum Neural Network Training Landscapes.
\newblock Nature Communications. 2018;9.

\bibitem{schuld2021}
Schuld M. Supervised quantum machine learning models are kernel methods; 2021.
\newblock Available from: \url{https://arxiv.org/abs/2101.11020}.

\bibitem{gyurik2023}
Gyurik C, Dunjko V. Exponential separations between classical and quantum learners; 2023.
\newblock Available from: \url{https://arxiv.org/abs/2306.16028}.

\bibitem{molteni2024}
Molteni R, Gyurik C, Dunjko V. Exponential quantum advantages in learning quantum observables from classical data; 2024.
\newblock Available from: \url{https://arxiv.org/abs/2405.02027}.

\bibitem{liu2021}
Liu Y, Arunachalam S, Temme K.
\newblock A rigorous and robust quantum speed-up in supervised machine learning.
\newblock Nature Physics. 2021;17:1013--1017.

\bibitem{hillier2015}
Hillier FS, Lieberman GJ.
\newblock Introduction to Operations Research.
\newblock McGraw-Hill; 2015.

\bibitem{pirnay2024}
Pirnay N, et~al.
\newblock An In-Principle Super-Polynomial Quantum Advantage for Approximating Combinatorial Optimization Problems via Computational Learning Theory.
\newblock Science Advances. 2024;10.

\bibitem{jordan2024}
Jordan SP, Shutty N, Wootters M, et~al.. Optimization by Decoded Quantum Interferometry; 2024.
\newblock Available from: \url{https://arxiv.org/abs/2408.08292}.

\bibitem{abbas2024}
Abbas A, Ambainis A, Augustino B, et~al.
\newblock Challenges and Opportunities in Quantum Optimization.
\newblock Nature Reviews Physics. 2024;.

\bibitem{glover2019}
Glover F, Kochenberger G, Du Y. A Tutorial on Formulating and Using QUBO Models; 2019.
\newblock Available from: \url{https://arxiv.org/abs/1811.11538}.

\bibitem{braine2021}
Braine L, Egger DJ, Glick J, et~al.
\newblock Quantum Algorithms for Mixed Binary Optimization Applied to Transaction Settlement.
\newblock IEEE Transactions on Quantum Engineering. 2021;2:1--8.

\bibitem{djidjev2023}
Djidjev HN. Quantum annealing with inequality constraints: the set cover problem; 2023.
\newblock Available from: \url{https://arxiv.org/abs/2302.11185}.

\bibitem{jones2020}
Jones EB, Kapit E, Chang CY, et~al.. On the Computational Viability of Quantum Optimization for PMU Placement; 2020.
\newblock Available from: \url{https://arxiv.org/abs/2001.04489}.

\bibitem{radvand2024}
Radvand T, Talebpour A. A Quantum Optimization Algorithm for Optimal Electric Vehicle Charging Station Placement for Intercity Trips; 2024.
\newblock Available from: \url{https://arxiv.org/abs/2410.16231}.

\bibitem{rao2023}
Rao PU, Sodhi B.
\newblock Hybrid quantum-classical solution for electric vehicle charger placement problem.
\newblock Soft Computing. 2023;27:13347--13363.

\bibitem{margarita2023}
Veshchezerova M, Somov M, Bertsche D, et~al.
\newblock A Hybrid Quantum-Classical Approach to the Electric Mobility Problem.
\newblock In: 2023 IEEE International Conference on Quantum Computing and Engineering (QCE). vol.~01; 2023. p. 636--641.

\bibitem{rozycki2023}
Różycki R, Józefowska J, Kurowski K, et~al.
\newblock A Quantum Approach to the Problem of Charging Electric Cars on a Motorway.
\newblock Energies. 2023;16(442).

\bibitem{aman2022}
Chandra A, Lalwani J, Jajodia B.
\newblock Towards an Optimal Hybrid Algorithm for EV Charging Stations Placement using Quantum Annealing and Genetic Algorithms.
\newblock In: 2022 International Conference on Trends in Quantum Computing and Emerging Business Technologies (TQCEBT); 2022. p. 1--6.

\bibitem{ehsani2024}
Ehsani D.
\newblock Quantum-Powered Battery Scheduling in Modern Distribution Grids [Ph.D. thesis].
\newblock University of Denver; 2024.
\newblock Electronic Theses and Dissertations, 2370.
\newblock Available from: \url{https://digitalcommons.du.edu/etd/2370}.

\bibitem{kea2023}
Kea K, Huot C, Han Y.
\newblock Leveraging Knapsack QAOA Approach for Optimal Electric Vehicle Charging.
\newblock IEEE Access. 2023;11:109964--109973.

\bibitem{marika2022}
Federer M, Müssig D, Klaiber S, et~al.
\newblock Application benchmark for quantum optimization on electromobility use case.
\newblock In: 2022 IEEE Vehicle Power and Propulsion Conference (VPPC); 2022. p. 1--6.

\bibitem{bucher2024}
Bucher D, Nüßlein J, O’Meara C, et~al.
\newblock Incentivizing Demand-Side Response Through Discount Scheduling Using Hybrid Quantum Optimization.
\newblock IEEE Transactions on Quantum Engineering. 2024;5:1–15.
\newblock Available from: \url{http://dx.doi.org/10.1109/TQE.2024.3407236}.

\bibitem{vanderLinde2023}
van~der Linde SG, van~der Schoot W, Phillipson F.
\newblock Efficient Quantum Solution for the Constrained Tactical Capacity Problem for Distributed Electricity Generation.
\newblock In: Krieger UR, Eichler G, Erfurth C, et~al., editors. Innovations for Community Services. Cham: Springer Nature Switzerland; 2023. p. 203--221.

\bibitem{silva2023}
Silva FFC, Carvalho PMS, Ferreira LAFM.
\newblock A quantum computing approach for minimum loss problems in electrical distribution networks.
\newblock Scientific Reports. 2023;13:10777.

\bibitem{kaseb2024}
Kaseb Z, Möller M, Vergara PP, et~al.
\newblock Power flow analysis using quantum and digital annealers: a discrete combinatorial optimization approach.
\newblock Scientific Reports. 2024;14:23216.

\bibitem{bauer2024}
Bauer N, Yeter-Aydeniz K, Kokkas E, et~al.. Solving Power Grid Optimization Problems with Rydberg Atoms; 2024.
\newblock Available from: \url{https://arxiv.org/abs/2404.11440}.

\bibitem{bucher2024evaluatingquantumoptimizationdynamic}
Bucher D, Porawski D, Wimmer B, et~al.. Evaluating Quantum Optimization for Dynamic Self-Reliant Community Detection; 2024.
\newblock Available from: \url{https://arxiv.org/abs/2407.06773}.

\bibitem{hang2022}
Jing H, Wang Y, Li Y, et~al.
\newblock Quantum Approximate Optimization Algorithm-Enabled DER Disturbance Analysis of Networked Microgrids.
\newblock In: 2022 IEEE Energy Conversion Congress and Exposition (ECCE); 2022. p. 1--5.

\bibitem{jaeho2019}
Choi J, Kim J.
\newblock A Tutorial on Quantum Approximate Optimization Algorithm (QAOA): Fundamentals and Applications.
\newblock In: 2019 International Conference on Information and Communication Technology Convergence (ICTC); 2019. p. 138--142.

\bibitem{yarkoni2022}
Yarkoni S, Raponi E, Bäck T, et~al.
\newblock Quantum annealing for industry applications: introduction and review.
\newblock Reports on Progress in Physics. 2022;85(10):104001.

\bibitem{saravanan2013}
Saravanan B, Das S, Sikri S, et~al.
\newblock A solution to the unit commitment problem—a review.
\newblock Frontiers in Energy. 2013;7:223--236.

\bibitem{xia2010}
Xia X, Elaiw AM.
\newblock Optimal dynamic economic dispatch of generation: A review.
\newblock Electric Power Systems Research. 2010;80(8):975--986.
\newblock Available from: \url{https://www.sciencedirect.com/science/article/pii/S0378779610000027}.

\bibitem{samantha2021}
Koretsky S, Gokhale P, Baker JM, et~al.
\newblock Adapting Quantum Approximation Optimization Algorithm (QAOA) for Unit Commitment.
\newblock In: 2021 IEEE International Conference on Quantum Computing and Engineering (QCE); 2021. p. 181--187.

\bibitem{nima2022}
Nikmehr N, Zhang P, Bragin MA.
\newblock Quantum Distributed Unit Commitment: An Application in Microgrids.
\newblock IEEE Transactions on Power Systems. 2022;37(5):3592--3603.

\bibitem{reza2023}
Mahroo R, Kargarian A.
\newblock Learning Infused Quantum-Classical Distributed Optimization Technique for Power Generation Scheduling.
\newblock IEEE Transactions on Quantum Engineering. 2023;4:1--14.

\bibitem{terrados2007}
Terrados J, Almonacid G, Hontoria L.
\newblock Regional energy planning through SWOT analysis and strategic planning tools.: Impact on renewables development.
\newblock Renewable and Sustainable Energy Reviews. 2007;11(6):1275--1287.
\newblock Available from: \url{https://www.sciencedirect.com/science/article/pii/S1364032105001000}.

\bibitem{seetharaman2016}
Seetharaman A, Sandanaraj LL, Moorthy MK, et~al.
\newblock Enterprise framework for renewable energy.
\newblock Renewable and Sustainable Energy Reviews. 2016;54:1368--1381.
\newblock Available from: \url{https://www.sciencedirect.com/science/article/pii/S136403211501206X}.

\bibitem{morteza2016}
Rahimiyan M, Baringo L.
\newblock Strategic Bidding for a Virtual Power Plant in the Day-Ahead and Real-Time Markets: A Price-Taker Robust Optimization Approach.
\newblock IEEE Transactions on Power Systems. 2016;31(4):2676--2687.

\bibitem{chen2019p2p}
Chen K, Lin J, Song Y.
\newblock Trading strategy optimization for a prosumer in continuous double auction-based peer-to-peer market: A prediction-integration model.
\newblock Applied Energy. 2019;242:1121--1133.
\newblock Available from: \url{https://www.sciencedirect.com/science/article/pii/S0306261919305045}.

\bibitem{yang2015}
Yang H, Yang S, Xu Y, et~al.
\newblock Electric Vehicle Route Optimization Considering Time-of-Use Electricity Price by Learnable Partheno-Genetic Algorithm.
\newblock IEEE Transactions on Smart Grid. 2015;6(2):657--666.

\bibitem{fang2011}
Fang K, Uhan N, Zhao F, et~al.
\newblock A new approach to scheduling in manufacturing for power consumption and carbon footprint reduction.
\newblock Journal of Manufacturing Systems. 2011;30(4):234--240.
\newblock Selected Papers of 39th North American Manufacturing Research Conference.
\newblock Available from: \url{https://www.sciencedirect.com/science/article/pii/S0278612511000690}.

\bibitem{anese2018}
Dall’Anese E, Guggilam SS, Simonetto A, et~al.
\newblock Optimal Regulation of Virtual Power Plants.
\newblock IEEE Transactions on Power Systems. 2018;33(2):1868--1881.

\bibitem{biamonte2017tn}
Biamonte J, Bergholm V. Tensor Networks in a Nutshell; 2017.
\newblock Available from: \url{https://arxiv.org/abs/1708.00006}.

\bibitem{wang2023tn}
Wang M, Pan Y, Xu Z, et~al.. Tensor Networks Meet Neural Networks: A Survey and Future Perspectives; 2023.
\newblock Available from: \url{https://arxiv.org/abs/2302.09019}.

\bibitem{tangpanitanon2022}
Tangpanitanon J, Mangkang C, Bhadola P, et~al.
\newblock Explainable natural language processing with matrix product states.
\newblock New Journal of Physics. 2022;24.

\bibitem{tang2019}
Tang E.
\newblock A quantum-inspired classical algorithm for recommendation systems.
\newblock In: Proceedings of the 51st Annual ACM SIGACT Symposium on Theory of Computing. ACM; 2019. Available from: \url{http://dx.doi.org/10.1145/3313276.3316310}.

\bibitem{arrazola2020}
Arrazola JM, Delgado A, Bardhan BR, et~al.
\newblock Quantum-inspired algorithms in practice.
\newblock {Quantum}. 2020 Aug;4:307.
\newblock Available from: \url{https://doi.org/10.22331/q-2020-08-13-307}.

\bibitem{gharehchopogh2023}
Gharehchopogh FS.
\newblock Quantum-inspired metaheuristic algorithms: comprehensive survey and classification.
\newblock Artificial Intelligence Review. 2023;56:5479--5543.

\bibitem{narayanan1996}
Narayanan A, Moore M.
\newblock Quantum-inspired genetic algorithms.
\newblock In: Proceedings of IEEE International Conference on Evolutionary Computation; 1996. p. 61--66.

\bibitem{ke2010}
Meng K, Wang HG, Dong Z, et~al.
\newblock Quantum-Inspired Particle Swarm Optimization for Valve-Point Economic Load Dispatch.
\newblock IEEE Transactions on Power Systems. 2010;25(1):215--222.

\bibitem{shor1996}
Shor PW.
\newblock Fault-tolerant quantum computation.
\newblock In: Proceedings of 37th Conference on Foundations of Computer Science; 1996. p. 56--65.

\bibitem{pan2024}
Pan F, Gu H, Kuang L, et~al.
\newblock Efficient Quantum Circuit Simulation by Tensor Network Methods on Modern GPUs.
\newblock ACM Transactions on Quantum Computing. 2024;Accepted on 31 August 2024.

\bibitem{monroe2021}
Monroe C, Campbell WC, Duan LM, et~al.
\newblock Programmable quantum simulations of spin systems with trapped ions.
\newblock Rev Mod Phys. 2021 Apr;93:025001.
\newblock Available from: \url{https://link.aps.org/doi/10.1103/RevModPhys.93.025001}.

\bibitem{wintersperger2023}
Wintersperger K, Dommert F, Ehmer T, et~al.
\newblock Neutral atom quantum computing hardware: performance and end-user perspective.
\newblock EPJ Quantum Technology. 2023;10:32.
\newblock Available from: \url{https://doi.org/10.1140/epjqt/s40507-023-00190-1}.

\bibitem{zhong2020}
Zhong HS, Wang H, Deng YH, et~al.
\newblock Quantum computational advantage using photons.
\newblock Science. 2020;370(6523):1460--1463.
\newblock Available from: \url{https://www.science.org/doi/abs/10.1126/science.abe8770}.

\bibitem{morten2020}
Kjaergaard M, Schwartz ME, Braumüller J, et~al.
\newblock Superconducting Qubits: Current State of Play [Journal Article].
\newblock Annual Review of Condensed Matter Physics. 2020;11(Volume 11, 2020):369--395.
\newblock Available from: \url{https://www.annualreviews.org/content/journals/10.1146/annurev-conmatphys-031119-050605}.

\end{thebibliography}

\end{document}